\documentclass[seceq]{ptptex}
\usepackage{graphicx}
\usepackage{epstopdf}
\usepackage{ulem}

\usepackage{dcolumn}
\usepackage{bm}

\newcommand{\nc}[1]{\newcommand{#1}}

\nc{\be}{\begin{eqnarray}}
\nc{\ee}{\end{eqnarray}}
\nc{\la}{\langle}
\nc{\ra}{\rangle}
\nc{\f}[2]{\frac{#1}{#2}}
\def\simge{\mathrel{%
       \rlap{\raise 0.511ex \hbox{$>$}}{\lower 0.511ex \hbox{$\sim$}}}}
\def\simle{\mathrel{
       \rlap{\raise 0.511ex \hbox{$<$}}{\lower 0.511ex \hbox{$\sim$}}}}
\nc{\Fsi}{F^{\bf 1}}
\nc{\Fad}{F^{\bf 8}}
\nc{\Fsy}{F^{\bf 6}}
\nc{\Fan}{F^{{\bf 3}^*}\!}
\nc{\Vsi}{V^{\bf 1}}
\nc{\Vad}{V^{\bf 8}}
\nc{\Vsy}{V^{\bf 6}}
\nc{\Van}{V^{{\bf 3}^*}\!}
\nc{\bx}{{\bf x}}
\nc{\by}{{\bf y}}

\title{
Application of fixed scale approach to static quark free energies  \\
in quenched and $2+1$ flavor lattice QCD \\
with improved Wilson quark action%
}

\author{
Yu \textsc{Maezawa}$^1$\footnote{Present address: Physics Department, Brookhaven National Laboratory, Upton, NY, 11973},
Takashi \textsc{Umeda}$^2$,
Sinya \textsc{Aoki}$^{3,4}$,
Shinji \textsc{Ejiri}$^5$, \\
Tetsuo \textsc{Hatsuda}$^{6,7}$, 
Kazuyuki \textsc{Kanaya}$^3$,
Hiroshi \textsc{Ohno}$^8$\footnote{Present address:
 Fakult\"{a}t f\"{u}r Physik, Universit\"{a}t Bielefeld, D-33501 Bielefeld, Germany.}
 \\(WHOT-QCD Collaboration)

}

\inst{
$^1$Mathematical Physics Lab., Nishina Center, RIKEN, Saitama 351-0198, Japan\\
$^2$Graduate School of Education, Hiroshima University, Hiroshima 739-8524, Japan\\
$^3$Faculty of Pure and Applied Sciences, University of Tsukuba, \\Tsukuba, Ibaraki 305-8571, Japan\\
$^4$Center for Computational Sciences, University of Tsukuba, \\Tsukuba, Ibaraki 305-8577, Japan\\
$^5$Graduate School of Science and Technology, Niigata University, \\Niigata 950-2181, Japan\\
$^6$Theoretical Research Division, Nishina Center, RIKEN, Saitama 351-0198, Japan\\
$^7$IPMU, The University of Tokyo, Kashiwa, Chiba 277-8583, Japan\\
$^8$Graduate School of Pure and Applied Sciences, University of Tsukuba, \\Tsukuba, Ibaraki 305-8571, Japan\\
}

\abst{
The free energies of static quarks and the Debye screening masses in the quark gluon plasma
 are studied using Polyakov-line correlation functions 
   in lattice QCD adopting the fixed-scale approach in which temperature is varied without 
 changing the spatial volume and the renormalization factors.
We calculate static-quark free energies in various color channels in the high temperature phase up to about 3.5 times the (pseudo-)critical temperature, 
performing lattice simulations both in quenched and $2+1$ flavor QCD. 
For the quenched simulations, we adopt the plaquette gauge action on anisotropic $20^3 \times N_t$ lattices with $N_t=8$--26 at the renormalized anisotropy $a_s / a_t \simeq 4$. 
For $2+1$ flavor QCD, we adopt the renormalization-group improved Iwasaki gluon action and the non-perturbatively $O(a)$-improved Wilson quark action on isotropic $32^3 \times N_t$ lattices with $N_t=4$--12 at $m_{\rm PS}/m_{\rm V} = 0.63$ (0.74) for the light (strange) flavors.
We find that the color-singlet free energies at high temperatures converge to the zero-temperature static-quark potential evaluated from the Wilson-loop at short distances. 
This is in accordance with the theoretical expectation that the short distance physics is insensitive to the temperature.
At long distances, the free energies approach to twice 
 the single-quark free energies, implying that the interaction between static quarks is fully screened.
We find that the static-quark free energies for various color channels turn out to be well described by the screened Coulomb form, and the color-channel dependence of the inter-quark interaction can be described by the kinetic Casimir factor inspired from the lowest order perturbation theory.
We also discuss comparison with a prediction of the thermal perturbation theory and flavor dependence of the screening masses.
}

\begin{document}

\maketitle

\section{Introduction}
\label{sec:intro}

At temperature around 150--200 MeV, the hadronic matter transforms into the quark-gluon plasma (QGP) by the QCD phase transition.
Recent experiments of relativistic heavy-ion collisions at RHIC and LHC
 make it possible to directly explore the properties of QGP \cite{YHM}.
Because the interaction is strong, lattice QCD simulations are indispensable in theoretical investigations of the QGP.
So far, most lattice simulations at finite temperature 
 have been performed by using staggered-type lattice quark actions 
  \cite{Bazavov:2010sb,Kaczmarek:2011zz,Borsanyi:2010bp,Borsanyi:2010cj,Endrodi:2011gv} with which the computational cost is less demanding.
However, the fourth-root procedure which is required to make a realistic quarks by the staggered-type actions is theoretically not fully established yet.
In this situation, it is of great importance to check the results by other actions such as the Wilson-type quark actions, which have solid theoretical basis.

A series of systematic studies of finite-temperature QCD with the $O(a)$-improved Wilson quark action was made by the CP-PACS Collaboration \cite{cp1,cp2}:
With two flavors of degenerate dynamical quarks, 
the phase structure, the pseudo-critical temperature, the O(4) critical scaling around the chiral transition, as well as the equation of state have been investigated.
The WHOT-QCD Collaboration is extending the project in various aspects --- to more detailed properties of QGP such as  the free energies of static quarks and the electric and magnetic screening masses \cite{Maezawa:2007fc,Ejiri:2009hq,Maezawa:2010vj}, 
to $2+1$ flavor QCD including the dynamical strange quark \cite{Umeda:2010ye,WEOS12}
and to finite densities \cite{WHOT10dense,WHOT12dense}. 
In this paper, we study static-quark free energies in various color channels as well as their Debye screening masses in QCD with $2+1$ flavors of improved Wilson quarks. 

The free energies for static quark-antiquark and two static quarks characterize inter-quark interactions in QGP,  
and their Debye screening masses describe the thermal fluctuation of quarks and gluons in QGP.
In a phenomenological model, they play important roles in the fate of heavy-quark bound states such as $J/\psi$ and $\Upsilon$ in QGP created at the center of heavy-ion collisions at RHIC and LHC \cite{Matsui-Satz,Satz:2008zc}.
On the lattice, the static-quark free energies are extracted from Polyakov-line correlation functions.
Studies in quenched QCD \cite{Kaczmarek:1999mm,Nakamura1,Kaczmarek:2004gv,Rothkopf:2011db}
and in full QCD with staggered-type quark actions \cite{Kaczmarek:2005ui,Doring,Fodor:2005qy,Doring:2005ih} or with Wilson-type quark actions \cite{Bornyakov:2004ii,Maezawa:2007fc,Ejiri:2009hq} have been made, 
and comparisons with analytic studies \cite{Laine:2006ns,Burnier:2009bk} have been attempted.
The studies with Wilson-type quarks were limited to the case of two-flavor QCD.
We now extend the study to $2+1$ flavors.

To reduce the high computational cost needed to simulate Wilson-type quarks, 
we adopt the fixed-scale approach \cite{Umeda:2008bd,Umeda:2010ye,WEOS12},
in which the temperature $T = 1/a N_t $ is varied by changing temporal lattice size $N_t$ with fixed values of the coupling parameters (and thus with a fixed value of the lattice spacing $a$).
Because the renormalization depends only on the coupling parameters, we can use a common zero-temperature simulation to obtain non-perturbative renormalization factors to renormalize observables at all temperatures.
We may even borrow existing configurations generated for zero-temperature studies.
We can thus reduce the costs for the zero-temperature simulations, which is quite demanding in finite-temperature studies.

In the fixed-scale approach, we vary $T$ without changing the system volume and lattice cut-off effects.
We can thus study the pure effects of $T$. 
We show later that this property of the fixed-scale approach benefits us also in a study of static-quark free energies.
On the other hand, results of the fixed-scale approach at very high temperatures, typically at $T \simge 1/3a$, suffer from lattice artifacts due to coarseness of the lattice to resolve thermal fluctuations
(see discussion in Sec.~\ref{sec:smq}).
We discuss this point later in this paper.

We first make test calculations of the static-quark free energies with the fixed-scale approach in quenched QCD.
We then study the case of $2+1$ flavor QCD adopting a non-perturbatively improved Wilson quark action and a renormalization-group improved Iwasaki gluon action 
at $m_{\rm PS}/m_{\rm V} = 0.63$ (0.74) for the light (strange) flavor, where $m_{\rm PS}$ $(m_{\rm V})$ is pseudo-scalar (vector) meson mass at $T=0$.
Comparison to the quenched results is made to identify the effects of dynamical quarks.
The lattice actions and the simulation parameters are summarized in Sec.~\ref{sec:2}. 
Results of the free energies of static quarks in various color channels are given in Sec.~\ref{sec:HQFE}.
We show that the color-singlet static-quark free energies at high temperatures universally converge to the zero-temperature static-quark potential evaluated from the Wilson-loop at short distances, 
whereas, at long distances, they approach to twice the single-quark free energy.
These are in accordance with theoretical expectations from the asymptotic freedom and the screening of colors in the high temperature deconfined phase.
Thanks to the fixed-scale approach, we directly confirm these properties.
The color-channel dependence of the free energies is also discussed.
In Sec.~\ref{sec:DSM}, the effective coupling $\alpha_{\rm eff}(T)$ and the Debye screening mass $m_D(T)$ are calculated by fitting the free energies with a screened Coulomb form. 
We discuss the effects of dynamical quarks and compare the results with a prediction of the thermal perturbation theory.
A summary is given in Sec.~\ref{sec:summary}.
In Appendix \ref{ap:eff_mass}, we discuss systematic errors in the Debye screening masses.

\section{Simulation parameters}
\label{sec:2}

\subsection{Quenched QCD}

In quenched QCD, we adopt configurations generated in Ref.~\citen{Umeda:2008bd} as the series ``a2'' 
on anisotropic $20^3 \times N_t$ lattices at the renormalized anisotropy $a_s / a_t \simeq 4$, 
where $a_s$ and $a_t$ are the lattice spacings in the spatial and temporal directions, respectively.
The anisotropic lattice enables us to study temperature dependence with fine
resolution in the fixed-scale approach since the resolution is proportional to $1/(N_t a_t)$.
This will be useful to compare the screening mass in the fixed-scale approach with
those in the previous fixed-$N_t$ approach in Sec.~\ref{sec:smq}.
In Ref.~\citen{Umeda:2008bd}, results obtained on isotropic lattice with a similar spatial lattice spacing are shown to be well consistent 
with those obtained on the anisotropic lattice.
The gauge action is the standard plaquette action with anisotropy,
\begin{eqnarray}
S &=& 
\beta \, \xi_0
\sum_x\sum_{i=1}^3
\left[ 1-\frac{1}{3}\mbox{Re}\mbox{Tr} U_{i4}(x) \right] + \frac{\beta}{\xi_0}
\sum_x\sum_{i>j;\, i,j=1}^3
\left[ 1-\frac{1}{3}\mbox{Re}\mbox{Tr} U_{ij}(x) \right]
\end{eqnarray}
where $U_{\mu\nu}(x)$ is the plaquette in the $\mu\nu$ plane
and $\beta$ and $\xi_0$ are the bare lattice gauge coupling and bare
anisotropy parameters. 
Tuning to the renormalized anisotropy $\xi=4$ as well as zero-temperature simulation on an anisotropic $20^3 \times 160$ lattice was done in Ref.~\citen{Matsufuru:2001cp}.

At our simulation point $\beta = 6.1$, the spatial cutoff scale is $a_s \simeq 0.097$ fm and the spatial length of the lattice is about 1.9 fm, where the lattice scale is set by the Sommer scale $r_0 = 0.5$ fm \cite{Matsufuru:2001cp}.
We use the configurations at $N_t=8$, 10, $\cdots$, 26 in the high temperature phase.
This range of $N_t$ corresponds to the temperature range $ 1.08 \simle T/T_{\rm c} \simle 3.50$, as summarized in Table~\ref{tab:paraQ},
where $T_{\rm c} \approx 290$ MeV is the critical temperature of the pure gauge theory. 
Up to a few million configurations were generated at each $N_t$ using the pseudo-heat-bath algorithm.
Although some configurations in the low temperature phase are also available, it turned out that the statistics is not high enough to extract static-quark free energies. 
We thus confine ourselves in the high temperature phase. 
Results for the equation of state with this action are given in Ref.~\citen{Umeda:2008bd}.

\subsection{$2+1$ flavor QCD with Wilson quarks}

For the simulations in $2+1$ flavor QCD, we employ the renormalization-group improved Iwasaki gluon action \cite{rg} coupled with 
$2+1$ flavors of the $O(a)$-improved Wilson quarks \cite{cl}:
\begin{eqnarray}
  S   &=& S_g + S_q \nonumber
, \\
  S_g &=& 
  -{\beta}\sum_x\left(
   c_0 \! \! \! \! \! \! \sum_{\mu<\nu;\,\mu,\nu=1}^{4}   \! \! \! \! \! \! W_{x,\mu\nu}^{1\times1} 
  +c_1 \! \! \! \! \! \! \sum_{\mu\ne\nu;\,\mu,\nu=1}^{4} \! \! \! \! \! \! W_{x,\mu\nu}^{1\times2}\right)
, \nonumber \\
  S_q &=& \sum_{f=u,d,s}\sum_{x,y}\bar{q}_x^f M_{x,y}^f q_y^f
, \nonumber
\end{eqnarray}
where $c_1=-0.331$, $c_0=1-8c_1$, and
 $W_{x,\mu\nu}^{n \times m}$ is the $n \times m$ rectangular Wilson loop.
The quark kernel is given by 
\begin{eqnarray}
 M_{x,y}^f &=& \delta_{x,y}
   - \kappa_f \sum_{\mu}\left\{(1-\gamma_{\mu})U_{x,\mu}\delta_{x+\hat{\mu},y} +(1+\gamma_{\mu})U_{x,\mu}^{\dagger}\delta_{x,y+\hat{\mu}}\right\}
 \nonumber \\ && \ \   -\delta_{x,y}\, \kappa_f \,{c_{SW}} \sum_{\mu<\nu}\sigma_{\mu\nu}F_{x,\mu\nu},
 \nonumber
\end{eqnarray}
where $\kappa_f$ is the hopping parameter for flavor $f$ and $F_{x,\mu\nu}$ is the lattice field strength given by the standard clover-shaped combination of gauge links.
The clover coefficient $c_{SW}$ is determined non-perturbatively as a function of $\beta$ \cite{Aoki:2005et}.

On isotropic $32^3 \times N_t$ lattices with $N_t=4$, 6, 8, 10, and 12, 
we perform simulations at $\beta = 2.05$, $(\kappa^{ud}, \kappa^s) = (0.1356, 0.1351)$ and $c_{SW} = 1.628$, which correspond to 
$m_{\rm PS}/m_{\rm V} = 0.6337(38)$ for light quarks and 
$m_{\rm PS}/m_{\rm V} = 0.7377(28)$ for the strange quark \cite{Ishikawa:2007nn}.
The lattice spacing is $a \simeq 0.071$ fm from the static-quark potential at $T=0$ (see Sec.~\ref{sec:HQFE}). 
The range $N_t=4$--12 correspond to $T \simeq 230$--700 MeV \cite{Umeda:2010ye}.
The results of Polyakov loop and its susceptibility as well as the equation of state suggest that 
the pseudo-critical temperature is around 190 MeV at our simulation point \cite{WEOS12}.

After thermalization, we generate configurations by the HMC+RHMC algorithm. 
The Metropolis test is performed every 0.5 trajectories for finite-temperature simulations.
We measure observables at every 5 trajectories.
The number of trajectories and corresponding temperatures for each $N_t$ are summarized in Table~\ref{tab:para}.
The statistical errors are estimated by the jackknife method with the bin-size of 20 trajectories \cite{WEOS12}.

\begin{table}[tbp]
 \begin{center}
 \caption{Temperature with respect to $T_c \simeq 290$ MeV for each $N_t$ in the quenched QCD simulations, where 
$T_c $ is the critical temperature of the pure gauge theory.
 For each temperature, we use 100--400 configurations to measure observables.
 $L$  is the Polyakov loop.}
 \label{tab:paraQ}
 {\renewcommand{\arraystretch}{1.1} \tabcolsep = 3mm
 \newcolumntype{a}{D{.}{.}{2}}
 \begin{tabular}{cccc}
 \hline\hline
 $N_t$ & $T/T_c$ & $\la L \ra$ \\
 \hline
 26 & 1.08 & 0.0542(5) \\
 24 & 1.17 & 0.0784(5) \\
 22 & 1.27 & 0.1051(3) \\
 20 & 1.40 & 0.1371(4) \\
 18 & 1.56 & 0.1767(4) \\
 16 & 1.75 & 0.2249(4) \\
 14 & 2.00 & 0.2847(3) \\
 12 & 2.33 & 0.3577(5) \\
 10 & 2.80 & 0.4462(2) \\
  8 & 3.50 & 0.5507(3) \\
 \hline\hline
 \end{tabular}}
 \end{center}
\end{table}

\begin{table}[tbp]
 \begin{center}
 \caption{Number of trajectories and temperatures
 for each $N_t$ in $2+1$ flavor QCD simulations.
Temperatures are estimated using the Sommer scale $r_0=0.5$ fm.
 $L$  is the Polyakov loop.}
 \label{tab:para}
 \begin{tabular}{cc}
 {\renewcommand{\arraystretch}{1.1} \tabcolsep = 3mm
 \newcolumntype{.}{D{.}{.}{2}}
 \begin{tabular}{cccl}
 \hline\hline
 \multicolumn{1}{c}{$N_t$} & 
 \multicolumn{1}{c}{Traj.} & 
 \multicolumn{1}{c}{$T$ (MeV)}&
 \multicolumn{1}{c}{$\la L \ra$}\\
 \hline
 12 & 6460 &   232 & 0.00491(\;\,6) \\ 
 10 & 3935 &   279 & 0.01470(11) \\
   8 & 2565 &   348 & 0.04072(12) \\ 
   6 & 2495 &   465 & 0.10981(11) \\ 
   4 & 3510 &   697 & 0.29185(\;\,7) \\ 
 \hline\hline
 \end{tabular}}
 \end{tabular}
 \end{center}
\end{table}

\section{Free energy of static quarks}
\label{sec:HQFE}

\begin{figure}[tbp]
  \begin{center}
    \includegraphics[width=85mm]{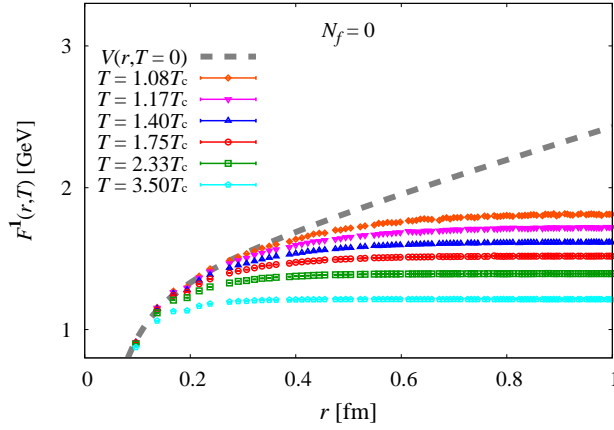} 
     \caption{Free energies of static quarks in color-singlet channel
    of quenched QCD at finite temperature.
    The dashed gray curve shows the static-quark potential $V(r)$ at $T=0$  
     calculated in Ref.~\citen{Matsufuru:2001cp}.
    The arrows on the right side denote 
    twice the single-quark free energy defined by
    $ 2 F_Q = - 2 T \ln \la {\rm Tr} \Omega \ra $.}
    \label{fig:HQFEQ}
  \end{center}
\end{figure}

\begin{figure}[tbp]
  \begin{center}
    \includegraphics[width=85mm]{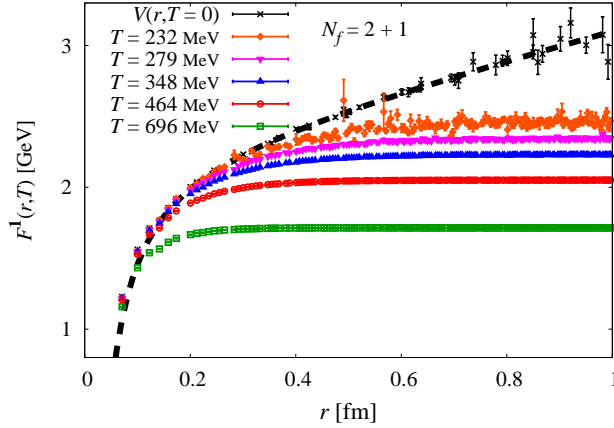}
    \caption{The same as Fig.~\ref{fig:HQFEQ} but for $2+1$ flavor QCD.
    The static-quark potential $V(r)$ at $T=0$ has been calculated
     by the CP-PACS and JLQCD Collaborations \cite{Ishikawa:2007nn}.
    The fit results of $V(r)$ by the Coulomb $+$ linear form are shown by the dashed gray curve.
    The arrows on the right side denote
    $ 2 F_Q = - 2 T \ln \la {\rm Tr} \Omega \ra $.}
    \label{fig:HQFE}
  \end{center}
\end{figure}

We now study the free energy of static quarks at zero and finite temperatures. 

\subsection{Static-quark potential at $T=0$}

At $T=0$, interaction between static quark ($Q$) and antiquark ($\bar{Q}$) can be described by the static-quark potential evaluated from the Wilson loop operator;
\be
V(r) = - \lim_{\ell \rightarrow \infty} \left[  \ell^{-1} \ln \left\langle W_{i4}^{r \times \ell} \right\rangle \right]
,
\ee
where $r$ is the distance between $Q$ and $\bar{Q}$.
It is  expected that the static-quark potential shows 
 the Coulomb-type behavior at short distances
 due to perturbative gluon exchange, while it takes 
 the linear behavior at long distances due to the confinement.
Therefore, its phenomenological form may be
 written as,
\be
 V(r) = - \frac{\alpha_0}{r} + \sigma r + V_0 
.
\label{eq:V}
\ee
The Sommer scale $r_0$ is defined as $r^2 dV/dr|_{r=r_0} = 1.65$.
We set the scale by a phenomenological value $r_0 = 0.5$ fm.

In quenched QCD, a fit to the simulation results obtained on the anisotropic $20^3 \times 160$ lattice with Eq.~(\ref{eq:V}) gives \cite{Matsufuru:2001cp} 
\begin{align}
(\alpha_0 a_s/a_t, \ \sigma a_s a_t, \ V_0 a_t , \ r_0 / a_s )
   = (0.068(2) , \ 0.0132(1) , \ 0.170(1) , \ 5.140(32) ) ,
\nonumber 
\end{align}
where the fitting range was chosen to be $1.73 \le r/a_s \le 10$.
Setting $r_0 = 0.5$ fm gives $1/a_s = 2.030 (13)$ GeV ($a_s \simeq 0.097$ fm) and $\sqrt{\sigma} \simeq 0.47$ GeV.
The fit result of the static-quark potential is shown by the dashed gray curve in Fig.~\ref{fig:HQFEQ}.

For $2+1$ flavor QCD, the static-quark potential was measured by 
 the CP-PACS and JLQCD Collaborations with the same simulation parameters
 on the lattice $N_s^3 \times N_t = 28^3 \times 56$ \cite{Ishikawa:2007nn}. 
Results are shown in Fig.~\ref{fig:HQFE} by the ``$\times$'' symbol.
Because the string-breaking effect is not manifest in the range of $r$ we study, 
we fit the data in the fit range of 2--$4\le r/a \le 14$  with Eq.~(\ref{eq:V}) to obtain
\begin{align}
(\alpha_0, \ \sigma a^2, \ V_0 a , \ r_0 / a )
   = (0.44(3) , \ 0.024(1) , \ 0.801(5) , \ 7.06(3) ) .
\nonumber 
\end{align}
We find $1/a = 2.79(1)$ GeV ($a \simeq 0.071$ fm) and $\sqrt{\sigma} \simeq 0.43$ GeV.
The fit result is shown in Fig.~\ref{fig:HQFE} by the dashed gray curve.
We find that the phenomenological form reproduces the lattice data well.

\subsection{Static-quark free energies in the high-temperature phase}

At $T>0$, we calculate the Polyakov loop defined as
\be
L &\equiv& {\rm Tr} \, \Omega(\bx) = {\rm Tr} \, \prod_{\tau=1}^{N_t} U_{(\bx,\tau),4}
,
\ee
where $U_{(\bx,\tau),4}$ is the link variable in the temporal direction. 
Results of the Polyakov loop expectation value in quenched\footnote{
In the quenched approximation of QCD, we take the real part of $Z(3)$-rotated Polyakov loop as the Polyakov loop expectation value
for the single quark free energy $F_Q = - T \ln \la L \ra$, 
because this $Z(3)$ branch is chosen when dynamical quarks are present.}
 and $2+1$ flavor QCD
 are summarized in Table~\ref{tab:paraQ} and \ref{tab:para}, respectively.
The Polyakov loop is related to a single-quark free energy defined as
 $F_Q = - T \ln \la L \ra$,  which expresses the free energy of a static quark
 in the thermal medium.

The inter-quark interaction at $T>0$ may be characterized 
 by the free energy of a static quark and an antiquark  separated by a distance $r$,
 i.e.\ by the Polyakov-line correlation function.
The correlation function between $Q$ and $\bar{Q}$ can be decomposed into irreducible representations of color singlet and octet
 in the color space, i.e. $ {\bf 3} \otimes {\bf 3}^\ast = {\bf 1} \oplus {\bf 8}$, whereas that between $Q$ and $Q$ can be decomposed into color antitriplet and octet,
  i.e. ${\bf 3} \otimes {\bf 3} = {\bf 3}^\ast \oplus {\bf 6}$.
Under the Coulomb gauge fixing, these correlation functions are given as \cite{Nadkarni1,Nakamura1},
\begin{align}
e^{-\Fsi (r,T)/T}
 &=
  \frac{1}{3} 
  \left\langle {\rm Tr} \! \left[ \Omega({\bf x}) \Omega^\dag ({\bf y}) \right] \right\rangle
, 
\label{eq:singlet}
\\
e^{-\Fad (r,T)/T}
 &=
 \frac{1}{8} 
 \left\langle {\rm Tr} \! \left[ \Omega({\bf x}) \right] {\rm Tr} \! \left[ \Omega^\dag ({\bf y}) \right] \right\rangle
 - \frac{1}{24} 
  \left\langle {\rm Tr} \! \left[ \Omega({\bf x}) \Omega^\dag ({\bf y}) \right] \right\rangle
 \label{eq:octet}
, \\
e^{-\Fsy (r,T)/T}
 &= 
  \frac{1}{12}
 \left\langle {\rm Tr} \! \left[ \Omega({\bf x}) \right] {\rm Tr} \! \left[ \Omega ({\bf y}) \right] \right\rangle
+ \frac{1}{12} 
  \left\langle {\rm Tr} \! \left[ \Omega({\bf x}) \Omega ({\bf y}) \right] \right\rangle
, \\
e^{-\Fan (r,T)/T}
 &= 
\frac{1}{6}
 \left\langle {\rm Tr} \! \left[ \Omega({\bf x}) \right] {\rm Tr} \! \left[ \Omega ({\bf y}) \right] \right\rangle
- \frac{1}{6}
  \left\langle {\rm Tr} \! \left[ \Omega({\bf x}) \Omega ({\bf y}) \right] \right\rangle
,
\label{eq:sextet}
\end{align}
where $r=|\bx - \by|$.

Figures \ref{fig:HQFEQ} and \ref{fig:HQFE} shows the singlet free energy $\Fsi(r,T)$ at different temperatures in quenched and in $2+1$ flavor QCD, respectively.
We find that, $\Fsi(r,T)$ at all temperatures converges to the zero-temperature static-quark potential $V(r)$ at short distances.
This is in accordance with the theoretical expectation that the short distance physics is insensitive to the temperature.
Note that, in the case of the conventional fixed-$N_t$ approach, because the simulation point is different for each $T$, $\Fsi(r,T)$ gets different renormalization depending on $T$. 
Therefore, we cannot directly compare the results of the bare free energies.
In early studies, this insensitivity was just assumed and used to adjust the constant term of $\Fsi(r,T)$ at each $T$ \cite{Kaczmarek:2004gv,Kaczmarek:2005ui}.
More recently, the renormalization factors are estimated by carrying out zero-temperature simulations at each finite temperature simulation point in the fixed-$N_t$ approach \cite{RBCB2008}.
In the present study with the fixed-scale approach, since the coupling parameters are common to all $T$'s, 
we can directly compare bare free energies at different $T$ and confirm the expected insensitivity at short distances without additional adjustments.

At large $r$, we find that $\Fsi(r,T)$ departs from $V(r)$ and eventually becomes flat due to the Debye screening.
In  Figs.~\ref{fig:HQFEQ} and \ref{fig:HQFE}, the asymptotic values of $\Fsi(r,T)$ at long distance are  compared with twice the single-quark free energy, $2F_Q(T)$,
denoted by the arrows on the right side of each plot.
We find that $\Fsi(r,T)$ converges to $2F_Q(T)$ quite accurately at long distances. 
This previously was observed also in two-flavor QCD with the fixed-$N_t$ approach \cite{Maezawa:2007fc}.

Comparing the results of the static-quark free energies in Figs.~\ref{fig:HQFEQ} and \ref{fig:HQFE}, we find no qualitative differences between quenched and $2+1$ flavor QCD.
In Sec.~\ref{sec:DSM}, we perform a more quantitative comparison in terms of the screening masses.

\begin{figure}[tbph]
  \begin{center}
  \begin{tabular}{cc}
    \includegraphics[width=68mm]{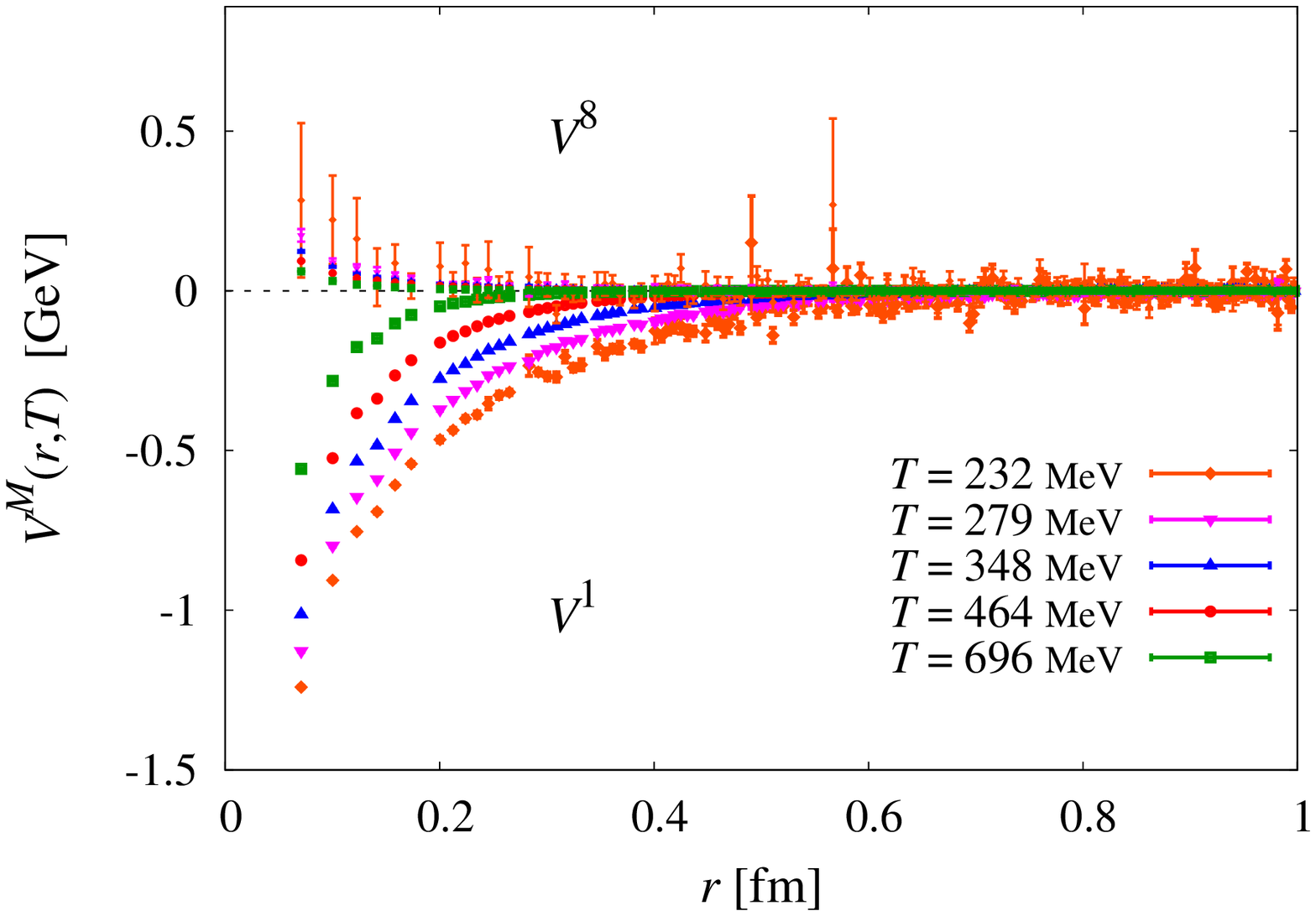} &
    \includegraphics[width=68mm]{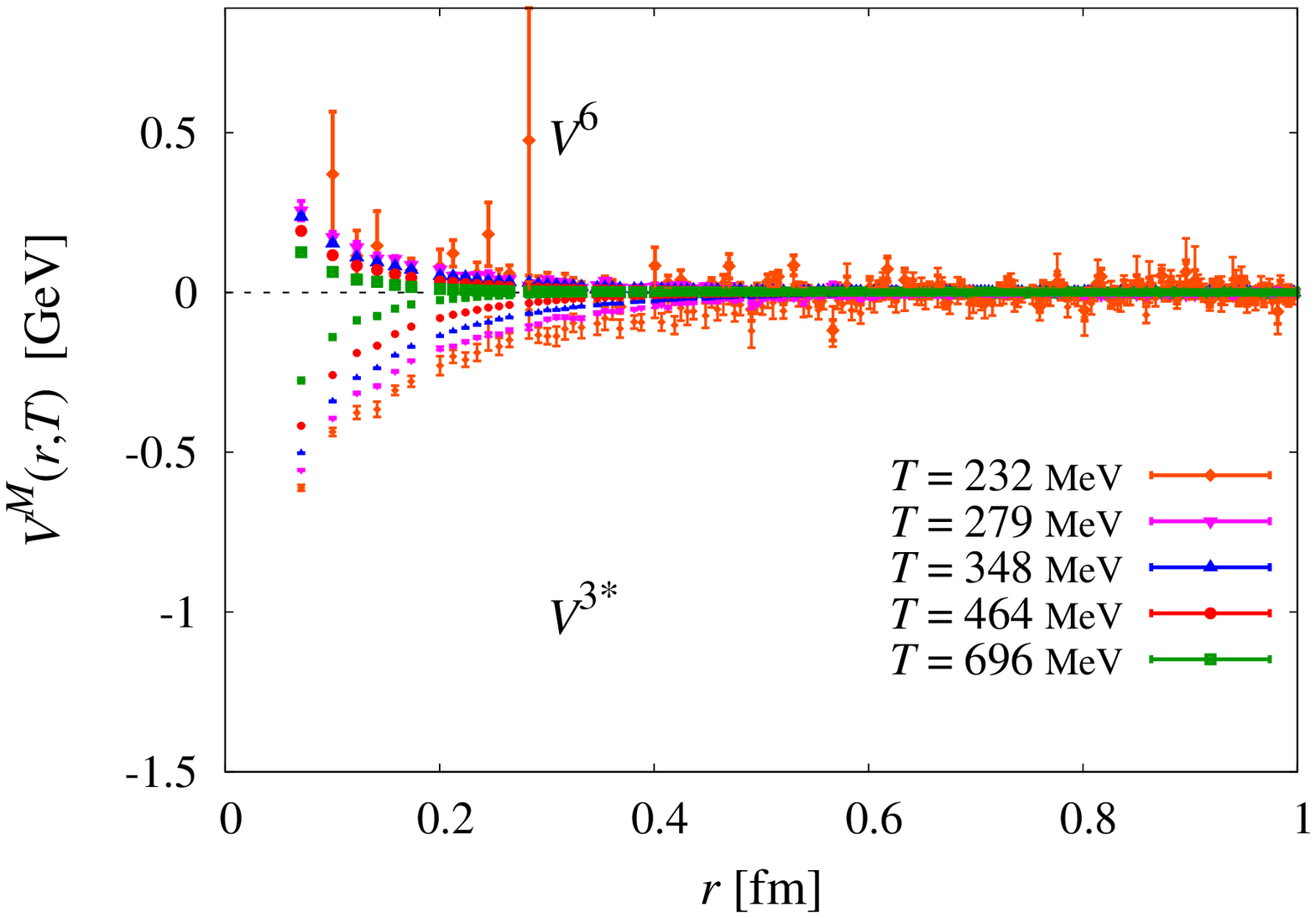}
  \end{tabular}
    \caption{Normalized free energies 
    for color-singlet and octet $Q\bar{Q}$ channels (left)
    and color-antitriplet and sextet $QQ$ channels (right)
    in $2+1$ flavor QCD with the improved Wilson quark action.}
    \label{fig:NFE}
  \end{center}
\end{figure}

 Let us now focus on the color-channel dependence by introducing normalized free energies
 $V^M(r,T) = F^M(r,T) - 2F_Q(T)$  (with $M={\bf 1}, {\bf 8}, {\bf 6}, {\bf 3}^*$), 
which are defined such that $V^M = 0$ at $r \rightarrow \infty$. 
 This is equivalent to define the free energies by 
 dividing the right-hand sides of 
 Eqs.~(\ref{eq:singlet})--(\ref{eq:sextet})
 by $\langle {\rm Tr} \Omega \rangle^2$, because $F^M(r,T)$ 
 for all channels  converges to $2F_Q(T)$ at $r \rightarrow \infty$.
Results in $2+1$ flavor QCD are shown in Fig.~\ref{fig:NFE} for color-singlet and octet $Q \bar{Q}$ 
 channels (left) and color-antitriplet and sextet $QQ$ channels (right).

We find that the inter-quark interactions, which are Debye-screened at long distances,  show ``attraction'' 
in the color singlet and antitriplet channels, while 
they show ``repulsion''  in the color octet and sextet channels, at intermediate and short distances.
These features have been reported also in quenched QCD \cite{Nakamura1} and in two-flavor QCD with improved Wilson quarks \cite{Maezawa:2007fc} in the fixed-$N_t$ approach.

\section{Screening masses}
\label{sec:DSM}

To study screening properties at long distances in each color-channel more quantitatively,
we fit the static-quark free energies by the screened Coulomb form
\begin{equation}
V^M (r,T) = C_{\! M} \, \frac{\alpha_{\rm eff}(T)}{r} \, e^{-m_D(T) \, r} ,
\label{eq:SCP}
\end{equation}
where $\alpha_{\rm eff}(T)$ and $m_D(T)$ are the effective running coupling and 
 the Debye screening mass, respectively.
Note that $V^M = 0$ at $r \rightarrow \infty$ with this ansatz, and thus we fit the normalized free energies by (\ref{eq:SCP}).

In the previous studies in quenched\cite{Nakamura1}
and two-flavor QCD\cite{Maezawa:2007fc,Ejiri:2009hq},
 the color-channel dependence of the static-quark free 
energies in the high temperature phase turned out to be well absorbed by the kinematical Casimir factors
$C_{\! M} \equiv \langle \sum_{a=1}^{8} t_1^a\cdot t_2^a \rangle_{\! M}$: 
\begin{eqnarray}
C_{\bf 1}       = -\frac{4}{3}, \hspace{5mm}
C_{\bf 8}       =  \frac{1}{6}, \hspace{5mm}
C_{\bf 6}       =  \frac{1}{3}, \hspace{5mm}
C_{{\bf 3}^*}   = -\frac{2}{3},
\label{eq:Casimir}
\end{eqnarray}
as inspired from the lowest order perturbation theory. 
This means that, by fixing the coefficient $C_{\! M}$ in (\ref{eq:SCP}) to the value given in (\ref{eq:Casimir}), we can well fit the normalized free energies with $\alpha_{\rm eff}(T)$ and $m_D(T)$ which are common to all color channels.
With our present data, we find that the normalized free energies in both quenched and $2+1$ flavor QCD have the same property.
Therefore, in the followings, we fix the coefficient $C_{\! M}$ to the kinematical Casimir factor (\ref{eq:Casimir}) to extract $\alpha_{\rm eff}(T)$ and $m_D(T)$.

\subsection{Screening mass in quenched QCD and comparison with fixed-$N_t$ approach}
\label{sec:smq}

\begin{figure}[tbp]
  \begin{center}
    \includegraphics[width=85mm]{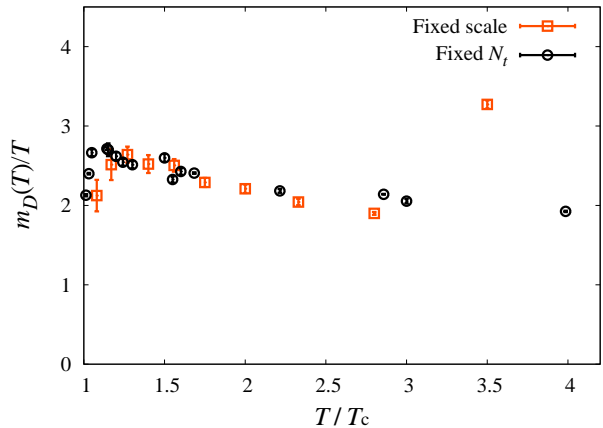}
    \caption{Debye screening mass for the singlet channel in quenched QCD. Open squares are the results obtained in the fixed-scale approach.
    For comparison, results obtained in the fixed-$N_t$ approach at $N_t=4$ \cite{Kaczmarek:2004gv} are also shown by open circles.}
    \label{fig:Nf0SM}
  \end{center}
\end{figure}

We first study the Debye screening masses in quenched QCD with the fixed-scale approach,
 and compare them with those obtained in a previous study in the fixed-$N_t$ approach.
To extract the Debye screening masses we
 fit the color-singlet free energies shown in Fig.~\ref{fig:HQFEQ} 
  with the screened Coulomb form (\ref{eq:SCP}).
The fit is performed by minimizing $\chi^2/N_{\rm DF}$ with the fit range of $1.0 \le rT \le 1.5$,
 and we obtain $\chi^2/N_{\rm DF} \sim 1$--2, except for highest temperature
 where we obtain $\chi^2/N_{\rm DF} =O(100)$.
An origin of the large $\chi^2 / N_{\rm DF}$ at high temperature is that we employ a fixed rage in $rT$ for the fits:
The number of data within the fit range decreases with increasing temperature.
Results of $m_D(T)$ are shown in Fig.~\ref{fig:Nf0SM}  as a function of temperature.
We also show the results of $m_D(T)$ obtained
 in the fixed-$N_t$ approach with the tree level Symanzik-improved
  gauge action on $N_s^3 \times N_t^{\rm (fix)} = 32^3 \times 4$ lattice \cite{Kaczmarek:2004gv}.
 We take a common value of $T_{\rm c}$  for both approaches.
We find that the temperature dependence of $m_D(T)$ in the fixed-scale approach
 is quantitatively consistent with that in the fixed-$N_t$ approach
 \cite{Kaczmarek:2004gv}
  except for the data at the highest temperature $T \simeq 3.5 T_{\rm c}$. 
A similar discrepancy between the two approaches at high $T$ has been reported also for the equation of state \cite{Umeda:2008bd}.
  
The discrepancy at the highest temperature should be regarded as a consequence of the limitation of the fixed-scale approach,
since the lattice spacing $a$ may be too coarse to resolve thermal fluctuations at very high $T$'s in the fixed-scale approach. 
We may estimate the range of applicability of the fixed-scale approach as follows:
Let us assume that a typical thermal wave length $\lambda$ at finite $T$ is given by $\lambda \sim 1/ E$ where $E$ is an average energy of massless particles.
We estimate $E \sim 3 T \zeta(4)/\zeta(3) \sim 2.7 T$ for the Bose-Einstein distribution 
and $E \sim 3 T \zeta(4)/\zeta(3) \times 7/6 \sim 3.15 T$ for the Fermi-Dirac distribution.
We thus obtain $\lambda \sim 1/(3T)$. 
On the present quenched lattice, $\lambda$ becomes as small as 0.65 fm at $T \simeq 3.5 T_c$, while our fixed lattice spacing in the quenched simulations is $a_s \simeq 0.97$ fm.
This is probably too coarse  to extract in-medium properties of thermal matter.

We should, therefore, take care with the data at temperatures $T \,\simge\, 1/(3a)$.
This is a potential drawback of the fixed-scale approach around the high temperature limit. 
In the fixed-$N_t$ approach, on the other hand, the lattice becomes finer as we increase $T$,
and thus the resolution for thermal fluctuation is kept fixed.
In the fixed-$N_t$ approach, however, the spatial volume may become quite small at high temperatures unless $N_s$ is increased simultaneously.

\subsection{Debye screening in $2+1$ flavor QCD}

We now study the Debye screening in $2+1$ flavor QCD by performing the screened Coulomb fits to the normalized free energies in all color channels.
Consulting the effective screening masses calculated in Appendix~\ref{ap:eff_mass}, we choose the interval $0.6 \le rT \le 1.25$ as the fit range.
The fit is performed by minimizing $\chi^2/N_{\rm DF}$.
Results of $\chi^2/N_{\rm DF}$ for each color-channel and temperature are summarized in Table~\ref{tab:chi2}. 
We find that the values of $\chi^2/N_{\rm DF}$ at the highest temperature ($T=697$ MeV) are quite large in contrast to other temperatures.
This is due to a fixed range in $rT$ discussed above.
In Appendix \ref{ap:eff_mass}, we also perform fit with a fixed range in $r/a$. 
We find that the results of this alternative fit are consistent with those of the present fit, and the systematic error estimated by the difference due to choice of the fit range is at most a few percents even at the highest temperature.

As the limiting temperature $T \sim 1/(3a)$ for the fixed-scale approach mentioned in the previous subsection, we have about 930 MeV for our lattice spacing $1/a = 2.79(1)$ GeV in $2+1$ flavor QCD.
Our highest temperature 697 MeV is lower than that.
However, since it is of the same order of this scale and $N_t=4$ there is rather small, we need to keep in mind possible influence of lattice artifacts around the highest temperature.

\begin{figure}[tbp]
  \begin{center}
  \begin{tabular}{cc}
    \includegraphics[width=68mm]{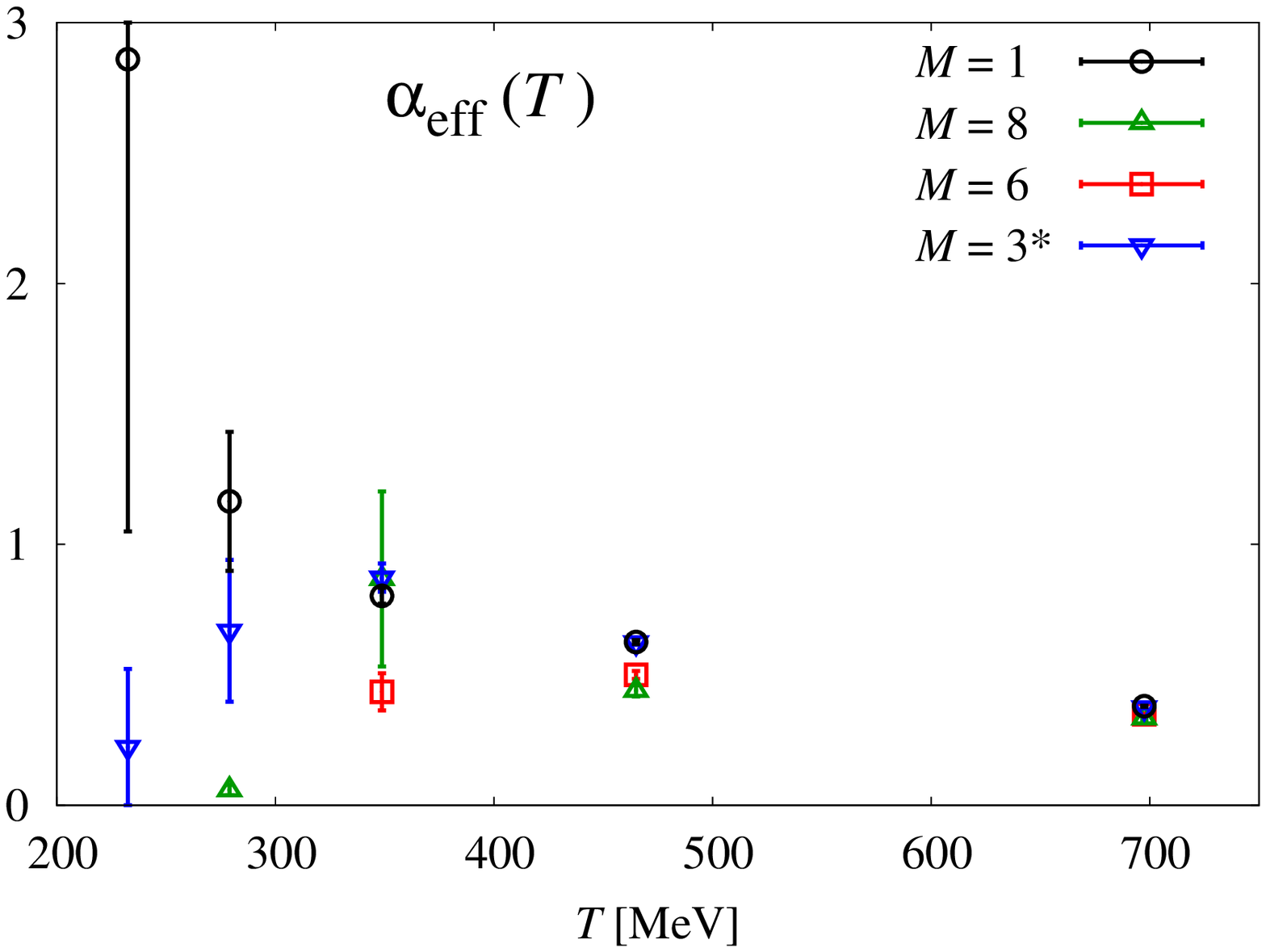} &
    \includegraphics[width=68mm]{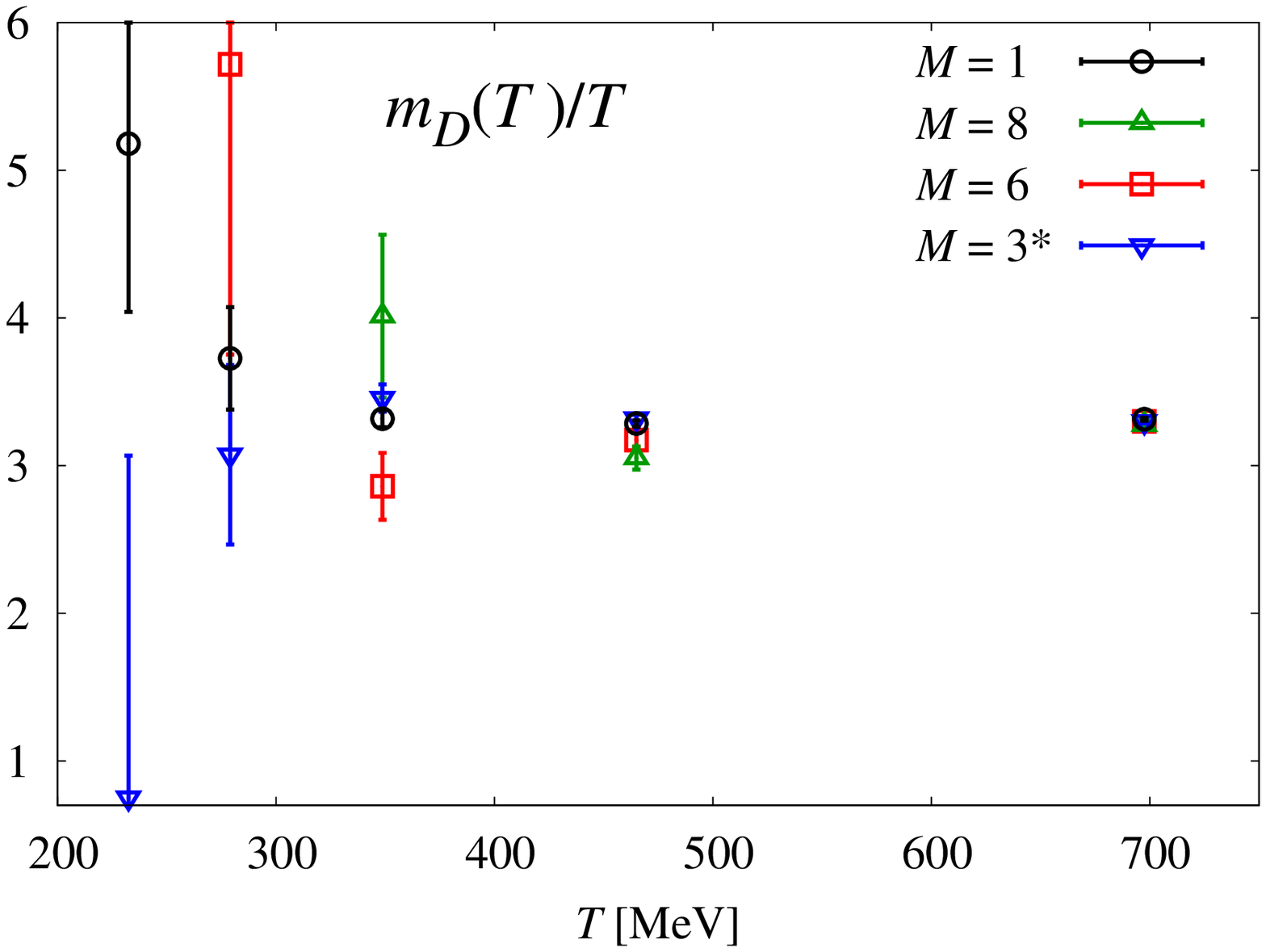}
    \end{tabular}
    \caption{Results of effective running coupling $\alpha_{\rm eff}$ (left)
    and Debye screening mass $m_D$ (right)
     for each color-channel as a function of temperature
     in $2+1$ flavor QCD.}
    \label{fig:DSM}
  \end{center}
\end{figure}

Results of $\alpha_{\rm eff}(T)$ and $m_D(T)$ are  summarized in Table~\ref{tab:ALMD} and plotted in Fig.~\ref{fig:DSM}.
We see that both $\alpha_{\rm eff}(T)$ and $m_D(T)$ are well independent on the color channel at $T > 348$ MeV, as implied by the Casimir dominance at high temperatures discussed at the beginning of this section.
This suggests that spatial correlations in different color-channels are dictated by the
same effective dynamics in the Coulomb gauge.

\begin{table}[tbp]
 \begin{center}
 \caption{Results of $\chi^2 / N_{\rm DF}$ in  $2+1$ flavor QCD
   for each color-channel and temperature.
  The fits at $T=232$ MeV for $M={\bf 8}$ and ${\bf 6}$
  are unstable since signals at these parameters are smaller than
   their statistical errors in the fitting interval.}
 \label{tab:chi2}
 \newcolumntype{.}{D{.}{.}{6}}
 \begin{tabular}{c....}
\hline\hline
 \multicolumn{1}{c}{$T$ [MeV]} &
 \multicolumn{1}{c} {$M={\bf 1}$} & 
 \multicolumn{1}{c} {${\bf 8}$} & 
 \multicolumn{1}{c} {${\bf 6}$} & 
 \multicolumn{1}{c}{${\bf 3}^*$} \\
\hline
\multicolumn{1}{c}{232} &
\multicolumn{1}{.} {1.42} & 
\multicolumn{1}{c} {$-$} & 
\multicolumn{1}{c} {$-$} & 
\multicolumn{1}{.} {2.43} \\
279 &  0.80 &  1.58 &  1.38 &  2.12 \\ 
348 &  1.05 &  1.18 &  1.82 &  2.53 \\ 
465 &  4.36 &  0.85 &  1.95 &  3.48 \\ 
697 & 69.36 &  8.96 & 22.92 & 48.77 \\
\hline\hline
 \end{tabular}
 \end{center}
\end{table}

\begin{table}[tbp]
 \begin{center}
 \caption{Results of $\alpha_{\rm eff} (T)$ and $m_D(T)$ in $2+1$ flavor QCD.
 The statistical errors are determined by a jackknife method with
 the bin-size of 20 trajectories.
 The second parentheses in the color-singlet channel
 show systematic errors due to the fit ranges,
 as explained in Appendix \ref{ap:eff_mass}.}
\label{tab:ALMD}
 \newcolumntype{.}{D{.}{.}{6}}
 {\renewcommand{\arraystretch}{1.0} \tabcolsep = 5mm
 \begin{tabular}{c|....}
 \hline\hline
 \multicolumn{1}{l|}{} &
 \multicolumn{4}{c}{$\alpha_{\rm eff}(T)$} \\
 \hline
 \multicolumn{1}{c|}{$T$ [MeV]} &
 \multicolumn{1}{c} {$M={\bf 1}$} & 
 \multicolumn{1}{c} {${\bf 8}$} & 
 \multicolumn{1}{c} {${\bf 6}$} & 
 \multicolumn{1}{c}{${\bf 3}^*$} \\
 \multicolumn{1}{c|}{232} &
 \multicolumn{1}{.} {2.86 (181) ( 37)} & 
 \multicolumn{1}{c} {$-$} & 
 \multicolumn{1}{c} {$-$} & 
  \multicolumn{1}{.}{0.22 ( 29)} \\
279 &  1.16 ( 26) ( 33) &  0.06 (  2) &  3.25 (424) &  0.67 ( 27) \\ 
348 &  0.80 (  2) (  0) &  0.87 ( 33) &  0.43 (  7) &  0.87 (  5) \\ 
465 &  0.63 (  1) ( 14) &  0.44 (  2) &  0.50 (  1) &  0.62 (  1) \\ 
697 &  0.38 (  1) ( 32) &  0.33 (  1) &  0.35 (  1) &  0.37 (  1) \\ 
 \hline
 \hline
 \multicolumn{1}{l|}{} &
 \multicolumn{4}{c} {$m_D(T)/T$} \\
 \hline
 \multicolumn{1}{c|}{$T$ [MeV]} &
 \multicolumn{1}{c} {$M={\bf 1}$} & 
 \multicolumn{1}{c} {${\bf 8}$} & 
 \multicolumn{1}{c} {${\bf 6}$} & 
 \multicolumn{1}{c}{${\bf 3}^*$} \\
 \multicolumn{1}{c|}{232} &
 \multicolumn{1}{.} { 5.18 (113) ( 27)} & 
 \multicolumn{1}{c} {$-$} & 
 \multicolumn{1}{c} {$-$} & 
 \multicolumn{1}{.} {0.75 (231)} \\
279 &  3.73 ( 34) ( 51) &  0.02 ( 36) &  5.72 (196) &  3.07 ( 60) \\ 
348 &  3.32 (  5) (  1) &  4.01 ( 55) &  2.86 ( 22) &  3.46 (  9) \\ 
465 &  3.29 (  2) ( 26) &  3.05 (  7) &  3.17 (  4) &  3.32 (  2) \\ 
697 &  3.31 (  1) ( 66) &  3.28 (  2) &  3.30 (  1) &  3.30 (  1) \\ 
 \hline\hline
 \end{tabular}}
 \end{center}
\end{table}

\begin{figure}[tbp]
  \begin{center}
 \begin{tabular}{cc}
    \includegraphics[width=68mm]{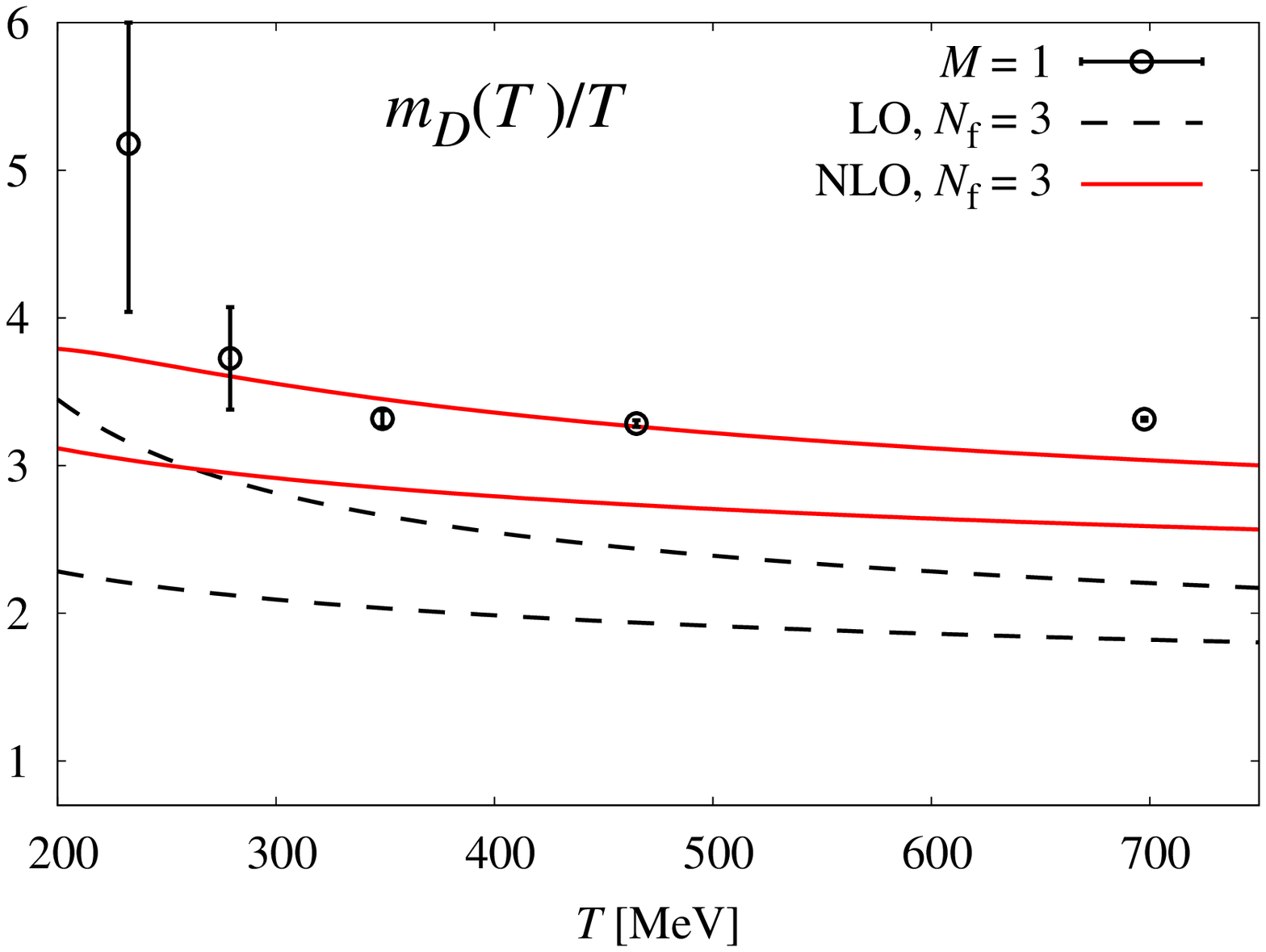} &
    \includegraphics[width=68mm]{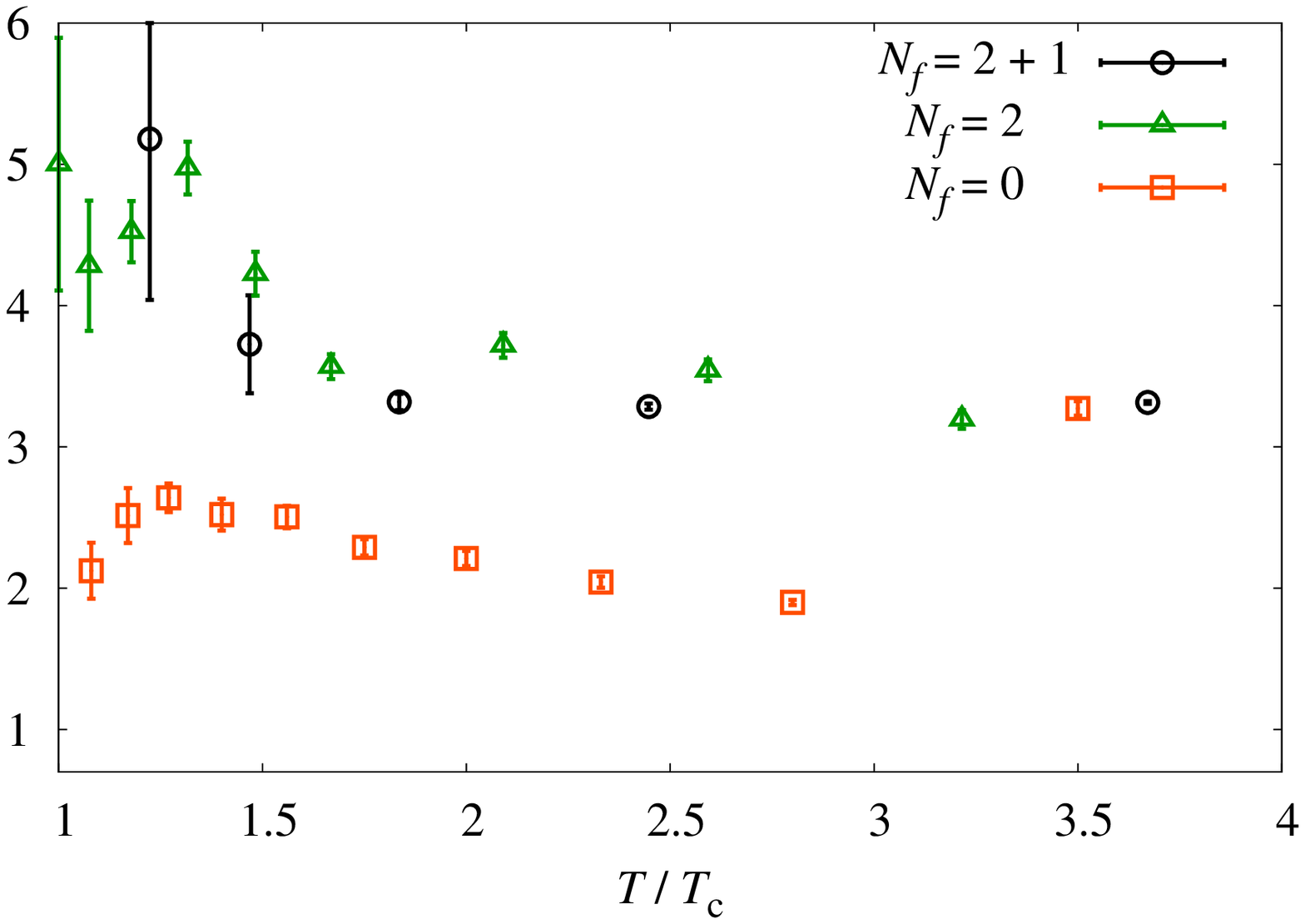}
    \end{tabular}
    \caption{(Left) The Debye screening mass on the lattice in the color-singlet channel together with that calculated 
    in the leading-order (LO) and next-to-leading-order (NLO) perturbation theory shown by
    dashed-black and solid-red lines, respectively. The bottom (top) line expresses a result at $\mu=\pi T$ ($3 \pi T$), 
    where $\mu$ is the renormalization point.
    (Right) Flavor dependence of the Debye screening masses. We assume the pseudo-critical temperature 
     for $2+1$-flavor QCD as $T_{\rm c} \sim 190$ MeV.}
    \label{fig:pQCD}
  \end{center}
\end{figure}

To make a qualitative comparison of the Debye masses between lattice QCD and thermal perturbation theory, in the left panel of Fig.~\ref{fig:pQCD}, 
we show our $N_f=2+1$ data of  $m_D(T)$ in the color-singlet channel,
together with the results of leading order (LO) and next-to-leading order (NLO) thermal perturbation theory in massless $N_f=3$ QCD. 
See Ref.~\citen{Maezawa:2007fc} for the LO and NLO formulae we adopt.
We find that NLO results and the lattice data are qualitatively consistent with each other, though there is always a concern about the convergence of thermal perturbation theory. 
The slight deviation at the highest temperature $T=696$ MeV may be due to the lattice artifact mentioned above.

Finally we study the flavor-dependence of the Debye screening masses.
In the right panel of Fig.~\ref{fig:pQCD}, we compare the results of $m_D(T)$ in the color-singlet channel
in $2+1$ flavor QCD, two-flavor QCD\cite{Maezawa:2007fc} and quenched (0-flavor) QCD as a function of $T/T_{\rm c}$, where $T_{\rm c}$ is the (pseudo) critical temperature. 
The two-flavor results are obtained in Ref.~\citen{Maezawa:2007fc} with improved Wilson quarks in the fixed-$N_t$ approach on a $16^3 \times 4$ lattice,  
where the light quark mass is $m_{\rm PS} / m_{\rm V} = 0.65$.
For $2+1$ flavor QCD, we assume $T_{\rm c} \simeq 190$ MeV \cite{WEOS12}.
Note that the two-flavor data are obtained with the fixed-$N_t$ approach with the cutoff scales changing from $a\sim0.26$ fm (at $T\sim T_{\rm c}$) to 0.07 fm (at $T\sim4T_{\rm c}$), while the $2+1$ flavor data in the fixed-scale approach are obtained at fixed $a \simeq 0.07$ fm.
The two-flavor data may suffer from cutoff effects at low temperatures.
Consulting the right panel of Fig.~\ref{fig:pQCD}, we find that magnitude of $m_D$ in $2+1$ flavor QCD is about 1.5 times as large as that in quenched QCD,
 except for the data at $T\simeq 3.5T_{\rm c}$ in quenched QCD which suffers from lattice artifacts 
 as discussed in the previous subsection. 
(The limiting temperature $T\sim 1/(3a)$ for the fixed scale approach is about $2.3 T_{\rm c}$ for the quenched data, and about $4.9 T_{\rm c}$ for the $2+1$ flavor data.)

This order of the magnitude for $m_D(T)$ in quenched and $2+1$ flavor QCD is roughly consistent with the prediction of the thermal perturbation theory in which the screening mass
 is predicted to be $m_D^{\rm LO} = \sqrt{1 + N_f/6\,} \, gT$ in the leading order where $g$ is the running coupling and the quark mass effects are neglected%
\footnote{
A more quantitative comparison may be attempted as follows:
Assuming one-loop running coupling $g^{-2} (\mu) = (4\pi)^{-2} \left( 11 - \frac{2}{3} N_f \right) \ln \left( \mu/\Lambda \right)^2$ and disregarding a small difference in $\Lambda$,\cite{Gockeler:2005rv}
we find $m_D(N_f=0) : m_D(N_f=2) : m_D(N_f=3) \sim 1 : 1.23 : 1.35$ in the massless limit,
to be compared with the lattice result $\sim 1 : 1.5 : 1.5$ shown in Fig.~\ref{fig:pQCD} (left) at finite quark masses.
 }.
On the other hand, we find that the magnitude of $m_D$ in $2+1$ flavor QCD 
  is almost comparable to that in two-flavor QCD at $m_{\rm PS} / m_{\rm V} = 0.65$ \cite{Maezawa:2007fc}, 
   which is approximately the same value with that in $2+1$ flavor QCD.
This suggests that the dynamical strange quark does not have large effect for the Debye screening mass in the temperature region we investigated.

\section{Summary}
\label{sec:summary}

We have studied static-quark free energies and Debye screening masses in finite temperature lattice QCD adopting the fixed-scale approach.
Unlike the conventional fixed-$N_t$ approach, the fixed-scale approach enables us to unambiguously estimate the temperature dependence of physical quantities without varying  the spatial volume and renormalization factors.
On the other hand, the fixed-scale approach may suffer from lattice artifacts at $T \simge 1/(3a)$ where  the thermal wave length of the plasma particles become comparable to the lattice spacing.

Performing simulations in quenched QCD and also in $2+1$ flavor QCD with non-perturbatively $O(a)$-improved Wilson quarks in the high temperature phase, 
we found that the static-quark free energies in the color-singlet channel converge to the zero temperature static-quark potential at short distances. 
This confirms the theoretical expectation that short distance physics is insensitive to the temperature.
At long distances, the free energies approach to twice 
 the single-quark free energies, implying that the interaction between static quarks is fully screened.
We also  confirmed that the inter-quark forces are 
 ``attractive'' in the color-singlet $Q \bar{Q}$ and antitriplet $QQ$ channels
 and ``repulsive'' in the color-octet $Q \bar{Q}$ and sextet $QQ$ channels in the case of $2+1$ flavor QCD. 
The static-quark free energies in these color-channels are well fitted by the screened Coulomb form with the kinetic Casimir factors, which are color-channel dependent, combined with color-channel independent effective coupling $\alpha_{\rm eff}(T)$ and Debye screening mass $m_D(T)$. 
These observations are in accordance with previous studies in quenched \cite{Nakamura1} and two-flavor QCD \cite{Maezawa:2007fc} in the fixed-$N_t$ approach.
Our result of $m_D(T)$ is qualitatively consistent with the next-to-leading order thermal
   perturbation theory except at the highest temperature corresponding to $N_t=4$.
Values of $m_D(T)$ in $2+1$ flavor QCD are larger than those in quenched QCD,
  which is again qualitatively consistent with a prediction of the thermal perturbation theory.
We also noted that $m_D(T)$ in $2+1$ flavor QCD are comparable to those in the two-flavor QCD obtained in the fixed-$N_t$ approach at a similar light quark masses \cite{Maezawa:2007fc}, implying that, at the quark masses of the present study, the influence of the dynamical strange quark is not large in the Debye screening mass in the temperature range we investigated.

\section*{Acknowledgements}

We thank the members of the CP-PACS and JLQCD Collaborations for 
providing us with their high-statistics 2+1 flavor QCD configurations at $T=0$, 
and also the authors of Ref.~\citen{Matsufuru:2001cp} for providing us with the results of their static-quark potential fit.
YM would like to thank Koichi Yazaki for useful discussions.
HO is supported by the Japan Society for the Promotion of Science for Young Scientists.
This work is in part supported 
by Grants-in-Aid of the Japanese Ministry
of Education, Culture, Sports, Science and Technology, 
(Nos.22740168, 
21340049  
20340047  
23540295).   
SA, SE and TH are supported in part by the Grant-in-Aid for Scientific Research on Innovative Areas (No.\ 2004: 20105001, 20105003, 2310576).
This work is in part supported also by the Large Scale Simulation Program of High Energy Accelerator Research Organization (KEK) Nos. 09/10-25 and 10-09, and the RIKEN Integrated Cluster of Cluster (RICC) facility.


\appendix

\section{Effective screening masses and systematic errors}
\label{ap:eff_mass}

\begin{figure}[tbp]
  \begin{center}
    \begin{tabular}{ccc}
    \includegraphics[width=44mm]{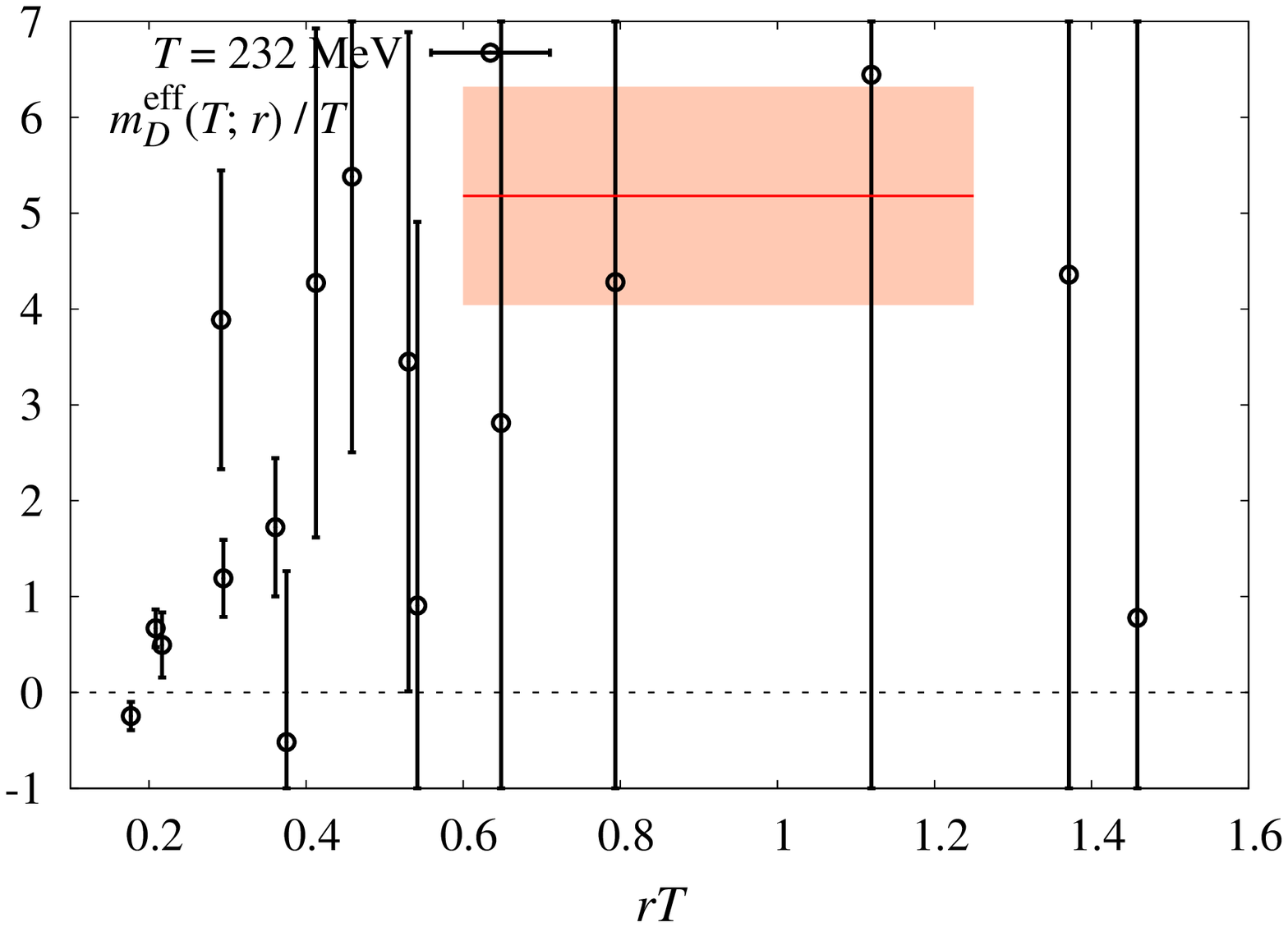} &
    \includegraphics[width=44mm]{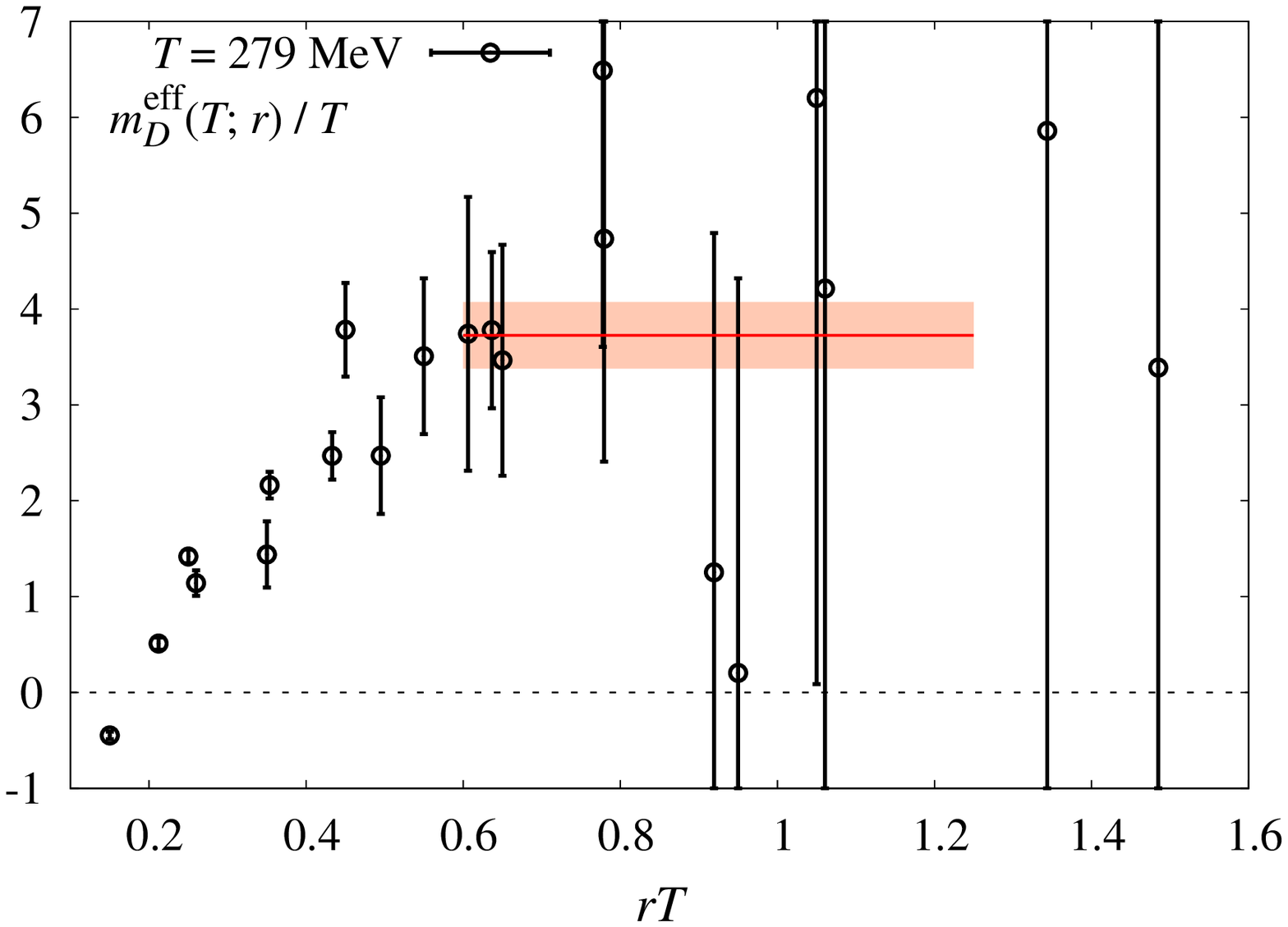} &
    \includegraphics[width=44mm]{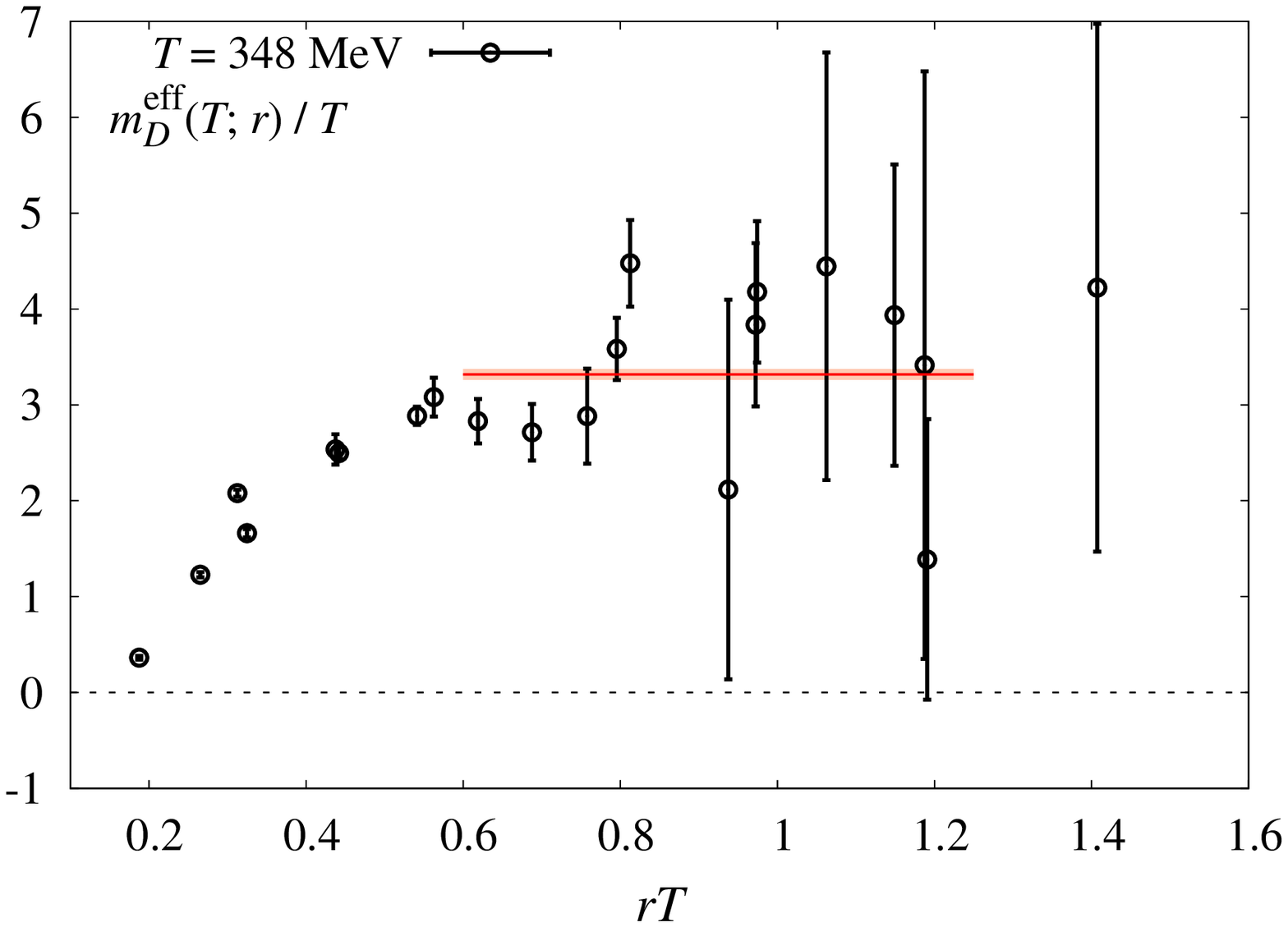} 
     \end{tabular}
    \begin{tabular}{cc}
    \includegraphics[width=44mm]{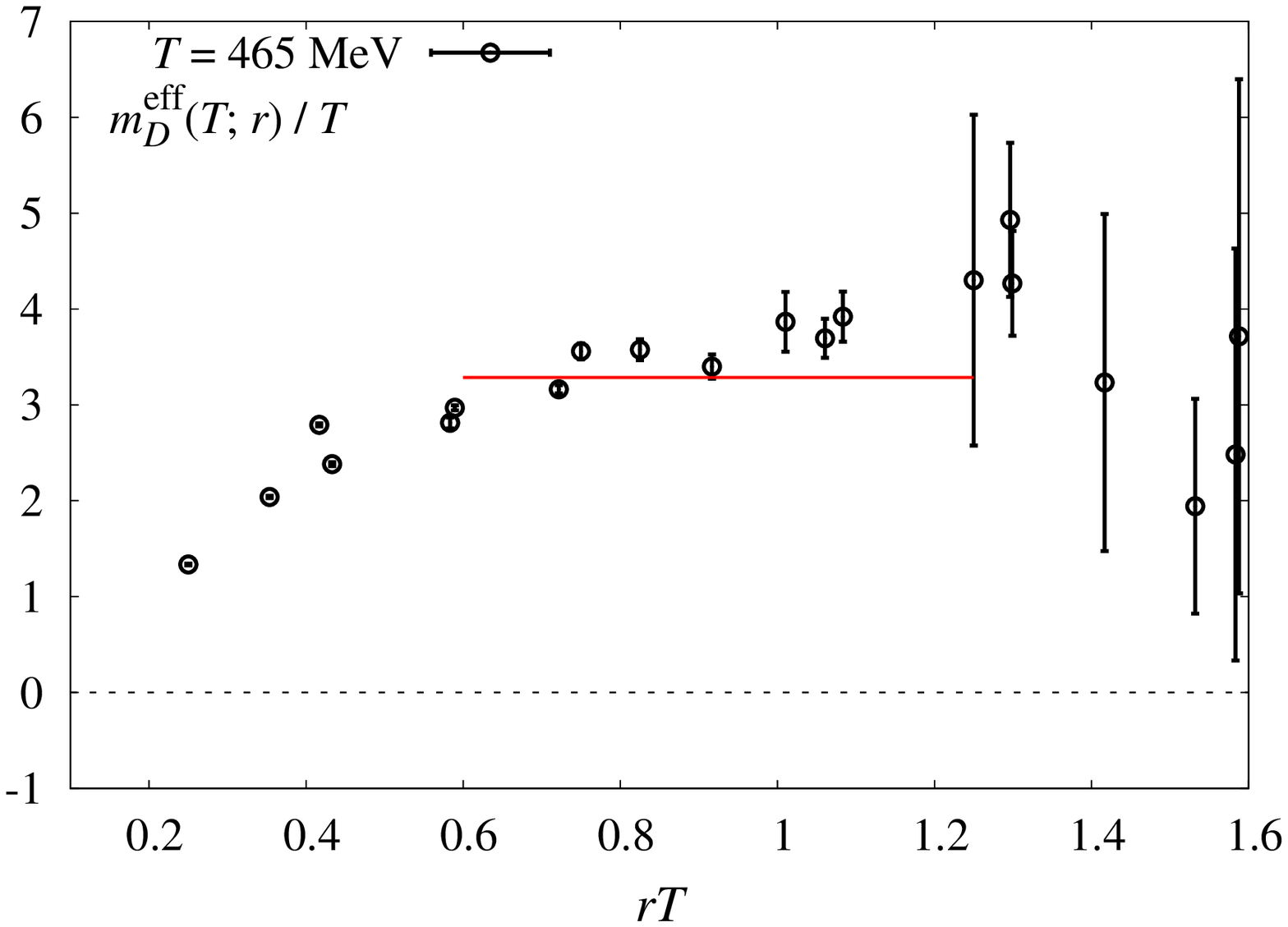} &
    \includegraphics[width=44mm]{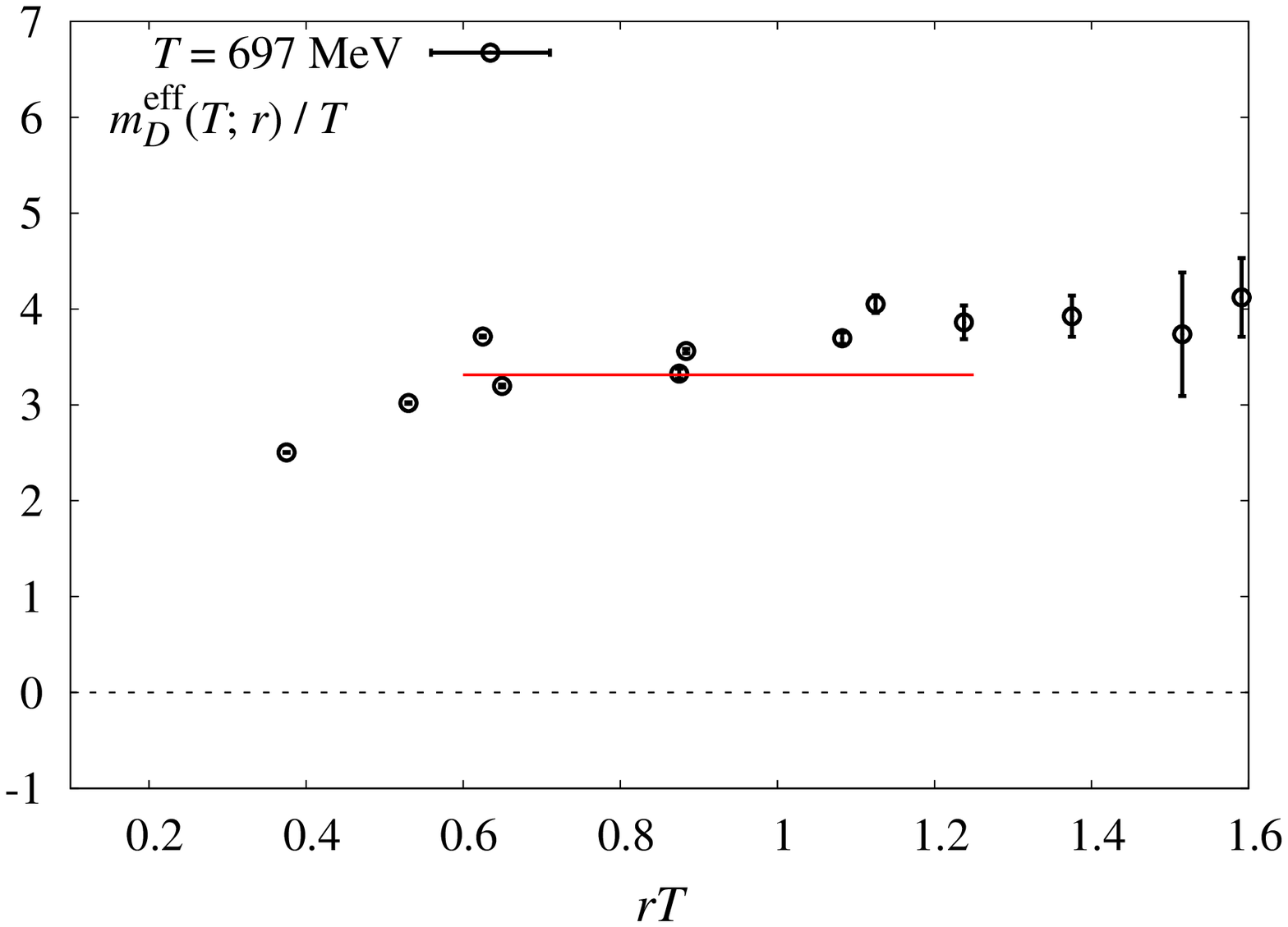}
       \end{tabular}
   \caption{Effective Debye screening mass at several temperatures as a function of $rT$.
    The shaded lines present the fit range with the fit results and statistical errors.}
    \label{fig:EDM}
  \end{center}
\end{figure}

In order to extract the screening mass $m_D$,
 we fit $V^M(r,T)$ by the screened Coulomb form, Eq.~(\ref{eq:SCP}).
To determine an appropriate fit range, we calculate the effective screening mass 
defined by 
\be
m_D^{\rm eff} (T;r) 
=
 \frac{1}{\Delta r} \log \frac{\Vsi(r)}{\Vsi(r+\Delta r)}
 - \frac{1}{\Delta r} \log \left[ 1 + \frac{\Delta r}{r} \right]
 \nonumber
\ee
for the color-singlet channel.
Figure \ref{fig:EDM} shows the results of effective screening masses at several temperatures 
 as a function of $rT$.
Since we find that the effective masses have a plateau at mid range of the distance,
 we choose the fit range to be $0.6 \le rT \le 1.25$.
The fit results of the screening masses shown as the lines on each figure
 reproduce the plateau well.
Results of $\chi^2/N_{\rm DF}$ for each color-channel 
 and temperature are summarized in Table \ref{tab:chi2}. 

Although Fig.~\ref{fig:EDM} suggests that the plateau region in $r$ is mainly determined by the thermal length $1/T$, we may attempt to fit the screening masses adopting a range independent of $T$.
We find that the range $4.5 \le r/a \le 8.0$ also leads to fits with acceptable $\chi^2/N_{\rm DF}$.
We estimate systematic errors due to a variation of the fit range by taking the difference of the fit results between $0.6 \le rT \le 1.25$ and $4.5 \le r/a \le 8.0$. 
The systematic errors thus estimated are given in the second parentheses of $\alpha_{\rm eff}(T)$ and $m_D(T)$ in Table \ref{tab:ALMD} for the color-singlet channel.
We find that the systematic errors are comparable to
 or smaller than statistical errors, except for the highest temperature.
The magnitude the systematic error at the highest temperature is, however, at most a few percents.


\end{document}